\documentclass[aps,prl,twocolumn,superscriptaddress,showpacs]{revtex4}

\usepackage{amssymb}
\usepackage{graphics}

\begin{document}


\title{Transport Phase Diagram of Homogeneously Disordered Superconducting Thin Films}



\author{Yize Li}
\affiliation{Department of Physics, University of Virginia, 
Charlottesville, VA, 22904, USA}

\author{Carlos L. Vicente}
\affiliation{Department of Physics, University of Virginia,
Charlottesville, VA, 22904, USA}
\affiliation{Universidad Puerto Rico, San Juan, Puerto Rico 00931, 
USA}

\author{Jongsoo Yoon}
\affiliation{Department of Physics, University of Virginia,
Charlottesville, VA, 22904, USA}


\date{\today}

\begin{abstract}
We have constructed a phase diagram in temperature-magnetic field-disorder 
space for homogeneously disordered superconducting tantalum thin films. 
Phases are phenomenologically identified by the nonlinear transport 
characteristics that are unique in each phase. The resulting phase diagram 
shows that a direct superconductor-insulator transition is prohibited at 
any disorder because the superconducting phase is completely surrounded by 
the intervening metallic phase. 
\end{abstract}


\maketitle

In two-dimensions (2D), a true superconducting state with zero electrical 
resistance is believed to exist only at zero temperature (T = 0). 
Increasing disorder in the system or applying magnetic fields (B) can disrupt the 
superconductivity, and eventually renders the system to an insulating 
state \cite{Finkelshtein, Larkin, Fisher1, Fisher2}. The transition 
between these two states in disordered films, however, is shrouded in 
mystery. While conventional theories, in either fermionic \cite{Finkelshtein, Larkin} or bosonic picture \cite{Larkin, 
Fisher1, Fisher2}, expect a direct superconductor-to-insulator transition, a mysterious metallic phase 
intervening the two phases is reported, notably in Ta \cite{Yoon1, Yoon2, 
Yoon3}, MoGe \cite{MK, EYKB}, and NbSi \cite{Aubin} films. The metallic 
state is usually identified by a drop in resistance followed by saturation to a finite value as $T\rightarrow0$ 
that is much smaller than the normal state resistance ($\rho_{n}$). The metallic state 
is also known to exhibit a nonlinear transport with $d^{2}V/dI^{2} > 0$ 
\cite{Yoon1, Yoon2, Yoon3}. This metallic nonlinear transport accompanies an extraordinarily long 
relaxation time, up to several seconds, and is demonstrated to arise from 
an intrinsic origin rather than to be a simple reflection of T-dependent 
resistivity via the combined effect of electron's failure to cool and 
the Joule heating \cite{Yoon1}. Three competing paradigms have been 
proposed to account for the emergence of the intervening metallic phase. 
The quantum vortex 
picture \cite{Galitski} describes the metallic phase as a Fermi-liquid of 
interacting vortices (vortex metal), the percolation paradigm \cite{SAK, 
GRT, DMA, SOK} describes the films as consisting of superconducting and normal puddles, and the phase 
glass model \cite{DP, WP} attributes the metallic transport to the 
coupling of the bosonic degrees of freedom to the excitations of the glassy phase.

So far, most of the experimental studies on the unexpected metallic phase 
are carried out in the lowest accessible temperatures. This is in part 
because the studies are more focused on the zero temperature ground state, 
and in part because of the lack of established experimental probes to 
characterize the metallic transport at elevated temperatures. In this 
Letter, we adopt nonlinear transport characteristics to identify the 
metallic phase and map the phase boundaries in the temperature-magnetic 
field plane. By combining such results from several films with different 
thicknesses representing different degree of disorder, we construct a 3D 
phase diagram in temperature-magnetic field-disorder space. The resulting 
phase diagram, which includes four different transport regimes, the 
superconducting, the metallic, the insulating, and the high temperature 
normal conducting phase, reveals two important features: 1) the 
superconducting phase is completely surrounded by the metallic phase 
prohibiting a direct superconductor-to-insulator transition at any 
disorder, and 2) the metallic phase appears even at zero field finite 
temperatures where the onset of superconductivity is usually understood in 
the framework of Kosterlitz-Thouless theory. In the rest of the paper, we 
describe how the phase boundaries are mapped and discuss the significance 
and implications of the resulting phase diagram.

\begin{table}
\caption{\label{table1}List of sample parameters: nominal film thickness 
t, normal state sheet resistance $\rho_{n}$ at 4.2 K, and the observed 
phases in each sample (S for superconducting phase, M for the metallic phase, I for 
the insulating phase, and N for the normal conducting phase). For samples 
exhibiting the superconducting phase, we list mean field $T_{c}$ at B = 0, 
the critical magnetic field $B_{c}$ as defined by the field at which the 
resistance reaches 90\% of the high field saturation value, and the 
correlation length calculated from $\xi = \sqrt{\Phi_{0} / 2\pi B_{c}}$ 
where $\Phi_{0}$ is the flux quantum.} 
\begin{ruledtabular} 
\begin{tabular}{lcccrccc}
Films & Batch & t(nm) & $\rho_{n}$(k$\Omega / \square$) & phase & 
$T_{c}$(K) & $B_{c}$(T) & $\xi$(nm) \\
Ta 1  & 1 & 5.6 & 1.42 & S,M,I,N & 0.65 & 0.82 & 20 \\
Ta 2  & 2 & 4.1 & 2.28 & S,M,I,N & 0.26 & 0.33 & 32 \\
Ta 3  & 3 & 2.5 & 3.04 &   M,I,N &      &      &    \\
Ta 4  & 3 & 2.4 & 3.34 &   M,I,N &      &      &    \\
Ta 5  & 3 & 2.3 & 3.54 &   M,I,N &      &      &    \\
Ta 6  & 3 & 2.2 & 3.88 &   M,I,N &      &      &    \\
Ta 7  & 3 & 2.1 & 4.20 &     I,N &      &      &    \\
Ta 8  & 3 & 2.0 & 4.98 &     I,N &      &      &    \\
Ta 9  & 3 & 1.0 & 6.20 &     I,N &      &      &    \\
Ta 10 & 4 & 2.5 & 6.24 &     I,N &      &      &    \\
Ta 11 & 5 & 2.5 & 8.00 &     I,N &      &      &    \\
\end{tabular}
\end{ruledtabular}
\end{table}

The results reported in this Letter are from dozens of Ta films of which 
sheet resistances at T = 4.2 K range from 0.07 - 700 k$\Omega / \square$. 
Parameters of 11 samples whose data are shown in this paper are summarized 
in Table I. The Ta films are dc sputter deposited on Si substrate after 
baking the chamber at $\sim 110^{\circ}$C for several days reaching a base pressure 
below $10^{-8}$ Torr. Sample deposition rate was $\sim$ 0.05 nm/s with an 
Ar pressure of $\sim$ 4 mTorr. Prior to the deposition the chamber and Ta source were cleaned by 
a pre-sputtering process. Using a rotatable substrate holder, up to 12 
films, each with a different thickness, can be grown without breaking the 
vacuum. Although there are noticeable batch to batch variations, the 
degree of disorder (evidence by the value of $\rho_{n}$) for films of the 
same batch increases monotonically with decreasing the nominal film thickness. 
All the samples were patterned into a Hall bar geometry (1 mm wide and 5 
mm long) for four point measurements using a shadow mask. Ta films 
prepared this way are known to be structurally amorphous and 
homogeneously disordered \cite{Yoon2}.

Shown in Fig.1(a) is a typical example of the T-dependence of 
resistance for a sample that undergoes a B-tuned 
superconductor-metal-insulator transition at the lowest temperature. At 
zero field (the bottom trace), the resistance at the lowest temperature is 
``immeasurably'' small and is in the superconducting phase, the traces 
shown as thin solid lines correspond to the metallic phase, and for the top two 
traces (dashed lines) the sign of $d\rho/dT$ is always negative which is 
usually taken as an insulating signature.

\begin{figure}
\includegraphics{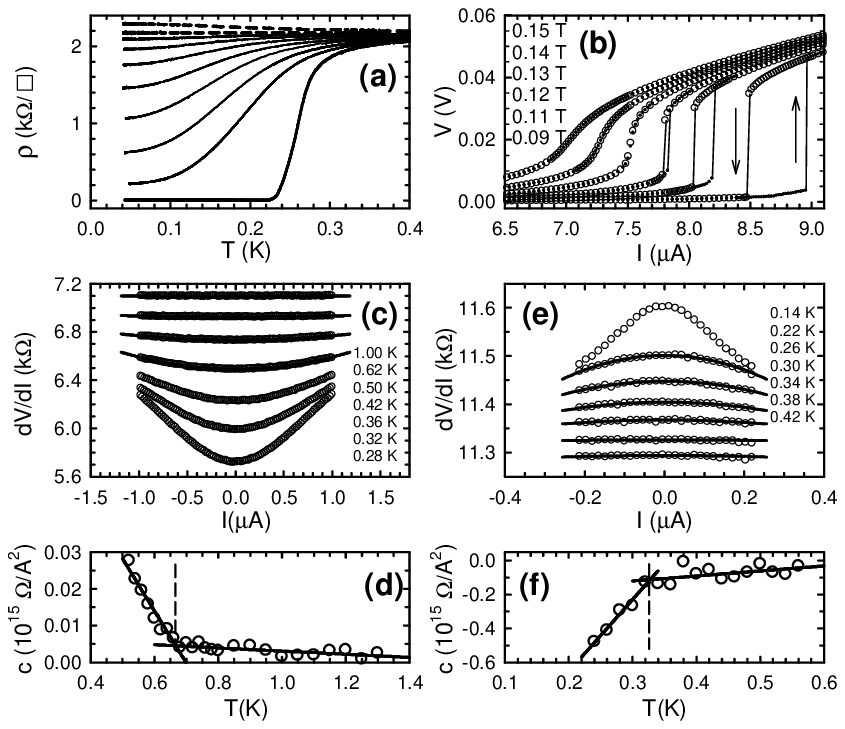}
\caption{\label{fig1}(a) Sheet resistance vs. T for Ta 2 at various 
magnetic fields in the range 0 -- 0.5 T. (b) I-V curves of Ta 1 at T = 0.1 
K at the indicated magnetic fields. Filled (open) circles are for current 
increasing (decreasing) branch. The solid lines are to indicate the 
electronic instabilities, and the arrows show the direction of the 
change. (c) dV/dI vs. I of Ta 1 at B  = 0.6 T at the indicated 
temperatures. The solid lines are the fit to the quadratic function. (d) 
The quadratic coefficients obtained from the fitting are plotted against 
temperature for Ta 1. The solid lines are a guide to an eye, and the 
dashed line corresponds to the metal-normal conductor phase boundary. (e) 
dV/dI vs. I for Ta 2 at B = 2.0 T at the indicated temperatures. Solid 
lines are fit to the quadratic function. (f) Quadratic coefficients vs. T 
for sample Ta 2. The dashed line corresponds to the insulator-normal 
conductor boundary.}
\end{figure}

We first discuss the phase identification at the lowest temperature by the 
nonlinear transport characteristics that are unique in each phase. As 
reported earlier \cite{Yoon1, Yoon2}, the superconducting phase exhibits a 
hysteretic I-V consisting of two electronic instabilities that appear as 
discontinuous voltage jumps occurring at different bias currents depending 
on whether the bias current is increasing or decreasing. Examples are 
shown in Fig.1(b). It is important to point out that 1) with 
approaching the instability the electronic relaxation time becomes 
increasingly longer, up to several seconds, and 2) the instabilities of 
the same characteristics are observed on films with normal state sheet 
resistances well below 100 $\Omega$. It has been already demonstrated 
\cite{Yoon1} that the instabilities are not due to electron's self-heating. Instead, 
the above two observations are consistent with the picture that the 
instabilities correspond to the vortex pinning-depinning phenomena arising 
from the competition between Lorentz driving force on the vortices and 
pinning force due to disorder \cite{Paltiel}. In this vortex picture all 
the vortices are expected to be pinned at zero temperature, which implies a 
realization of zero resistance state, or superconducting state, at zero 
temperature. With this reasoning, we judge the samples exhibiting 
hysteretic I-V's to be in their superconducting phase. With increasing B 
the hysteresis systematically evolves into a smooth and reversible curve 
characterized by $d^{2}V/dI^{2} > 0$ as the system is driven into the 
metallic phase. With further increasing B, eventually the system becomes an 
insulator with $d\rho/dT < 0$, where $d^{2}V/dI^{2}$ is always negative. 
The origins of the nonlinear transport in the metallic and the insulating phase are not 
known. The metallic nonlinear transport might be caused by current-induced 
depinning of vortices as suggested by the accompanying extraordinarily 
long relaxation time, and the insulating nonlinear transport by a similar 
effect of current on localized Cooper pairs. Nevertheless, we take these 
nonlinear transport characteristics as a phenomenological indicator to 
distinguish the phases.

The phase boundaries are mapped by following the evolution of the 
nonlinear transport with changing T or B. The superconductor-metal 
boundary is determined as the temperature or field at which the hysteresis 
in I-V curve first disappears.  Shown in Fig.1(b) is an example of 
B-driven evolution of the hysteretic I-V in the superconducting phase. 
With increasing B, the hysteresis becomes progressively smaller and 
disappears at B $\approx$ 0.13 T, which we identify as a 
superconductor-metal boundary. The T-driven evolution is similar as 
reported in Ref. \cite{Yoon1}.

In the metallic phase, it is convenient to analyze the nonlinear transport 
in terms of differential resistance. As shown in Fig.1(c), the 
nonlinearity progressively becomes weaker with increasing temperature and 
eventually the transport becomes linear. The linear transport at high 
temperatures is natural because at sufficiently high temperatures the 
system should be in the normal conducting state where the transport is 
linear. In order to quantitatively identify the phase boundary, we fit the 
dV/dI vs. I trace to a quadratic function $dV/dI = cI^{2} + \rho_{0}$, 
where c is the quadratic coefficient and $\rho_{0}$ is the sample's sheet 
resistance in the zero bias current limit. When the nonlinearity is relatively weak, the fitting 
is quite reasonable as shown by the solid curves for the top four traces 
in Fig.1(c). In Fig.1(d) the quadratic coefficients obtained from 
the least squared error fittings are plotted as a function of temperature. 
This plot shows that the behavior of the quadratic coefficients changes 
abruptly at a well defined temperature: At low temperatures (below the 
dashed line) the coefficients are rather strongly temperature dependent. 
At high temperatures (above the dashed line) the coefficients are almost 
temperature independent and the values are close to zero, implying the 
transport is practically linear. We identify the boundary, the dashed line 
in Fig.1(d), as the phase boundary separating the metallic phase and 
the high temperature normal conducting phase.

Figure 1(e) shows a similar evolution of the nonlinear transport in the 
insulating phase. The same quadratic function is used to fit the data. 
When the nonlinearity is relatively weak the fitting is very reasonable 
[solid lines in Fig.1(e)]. Again, as shown in Fig.1(f), there is a 
well defined temperature across which the behavior of the quadratic 
coefficients is distinctly different. We identify the low temperature 
regime (below the dashed line) where the T-dependence of the coefficients 
is relatively strong as the insulating phase, and the high temperature 
regime where the coefficients are almost T-independent and almost zero as 
the high temperature normal conducting phase. 

\begin{figure}
\includegraphics{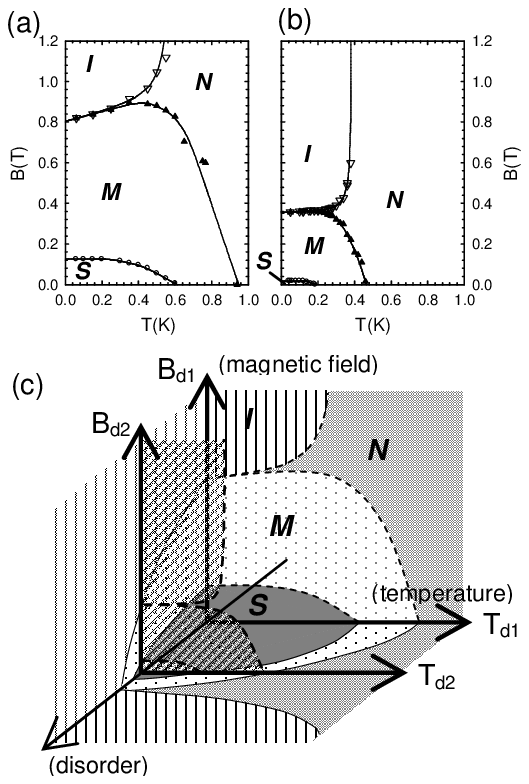}
\caption{\label{fig2}(a) Phase diagram of Ta 1 in the B-T plane. Symbols
are experimental data and solid lines are to guide an eye. The
superconducting phase is marked by S, the metallic phase by M, the
insulating phase by I, and the normal conducting phase by N. (b) Phase
diagram of Ta 2, which is more disordered than Ta 1. (c) The topology of
the 3D phase diagram in T-B-disorder space is shown. Disorder is
represented by the normal state sheet resistance and the axis only
indicates the direction of increasing disorder. The illustrated phase
diagram does not include the clean limit. The diagram shown in (a) for Ta
1 can be considered as a cross-sectional view on the plane of 
$T_{d1}-B_{d1}$, and the diagram for Ta 2 can be considered as a 
cross-sectional view on the plane of $T_{d2}-B_{d2}$.}
\end{figure}

The phase boundaries determined as described above are shown in Fig.2(a) and (b) for two samples, Ta 1 and Ta 2. Both samples are relatively 
weakly disordered and all four phases appear. The superconducting phase is 
marked by S, the metallic phase by M, the insulating phase by I, and the 
high temperature normal conducting phase by N. As represented by their 
normal state sheet resistances Ta 2 is more disordered than Ta 1. The 
results from these two samples capture the main feature that both the 
superconducting and the metallic phase shrink to lower temperature and 
lower field regime with increasing disorder and the insulating phase 
extends to lower fields. This tendency was clear in about a dozen other 
samples we investigated, although their data are not shown in this paper. 
This tendency is depicted in the 3D phase diagram in the T-B-disorder 
space in Fig.2(c), where the phase diagram of Ta 1 corresponds to a 
cross-sectional view in the plane of $T_{d1}-B_{d1}$ and the phase diagram 
of Ta 2 corresponds to a cross-sectional view at a higher disorder, 
$T_{d2}-B_{d2}$ plane.

\begin{figure}
\includegraphics{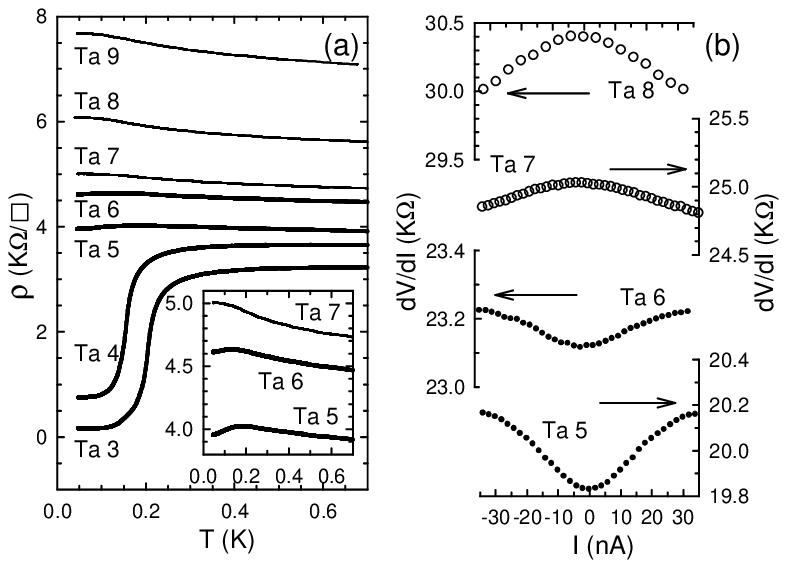}
\caption{\label{fig3}(a) T-dependence of resistance of samples Ta 3 -- 9 
at B = 0 is shown. All these samples are grown in a single batch. (inset) A 
blow-up view of resistance behavior at low temperatures for samples Ta 5 
-- 7. (b) Differential resistance (dV/dI) vs. I for samples Ta 5 -- 8. The arrows indicate the corresponding scale for each trace.}
\end{figure}

With further increasing disorder, we reach the regime where the 
superconducting phase no longer exists even at T = 0 and B = 0. This 
disorder regime appears as a gap between the superconducting and 
insulating phase along the disorder axis in Fig.2(c). In this disorder 
regime, the metallic phase occupies the low T and low B region in the 
phase diagram. Examples of such a metallic behavior at B = 0 is shown in 
Fig.3(a) as thick solid lines. The bottom two traces exhibit a rather 
steep drop in resistance followed by saturation to a finite value as 
$T\rightarrow0$. Samples Ta 5 and Ta 6 show only a weak T-dependence. 
However, as shown in the inset their slopes are positive at low temperatures and their 
nonlinear transport is characterized by $d^{2}V/dI^{2} > 0$ [filled 
circles in Fig.3(b)]. We emphasize that the metallic behaviors of these 
samples observed under B = 0 are indistinguishable from the B-induced metallic 
behaviors of samples of weak disorder such as Ta 1 or Ta 2. For even more 
disordered samples (samples Ta 7 -- 9), $d\rho/dT$ is always negative with 
nonlinear transport of $d^{2}V/dI^{2} < 0$ [open circles in Fig.3(b)], 
which are the insulating characteristics. In this high disorder regime where 
only the insulating phase appears at low temperatures, the boundary of the 
insulating phase grows to a higher temperature with increasing disorder. 
This tendency is shown in Fig.4(a) -- (c). 

\begin{figure}
\includegraphics{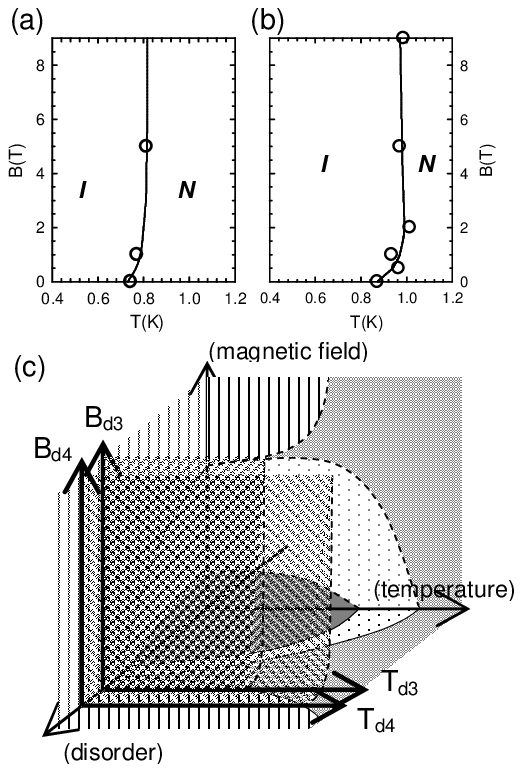}
\caption{\label{fig4}(a) Phase diagram of Ta 10 in the T-B plane. 
Insulating phase is marked by I and the normal conducting phase by N. 
Solid lines are to guide an eye. (b) Phase diagram of Ta 11 in the T-B 
plane. Ta 11 is more disordered than Ta 10 by the measure of $\rho_{n}$. 
(c) The same 3D phase diagram shown in Fig.2(c), but includes the 
corresponding cross-sectional views for Ta 10 ($T_{d3}-B_{d3}$ plane) and Ta 11 
($T_{d4}-B_{d4}$ plane). The hatched area in the $T_{d3}-B_{d3}$ 
($T_{d4}-B_{d4}$) plane corresponds to the insulating phase of Ta 10 (Ta 11).}
\end{figure}

With the presence of the B = 0 metallic phase tuned by disorder, the 
superconducting phase is completely surrounded by the metallic phase 
prohibiting a direct superconductor-insulator transition at any disorder. 
This is a fundamental difference from the phase diagram proposed earlier 
\cite{SBK}. Authors of Ref. \cite{SBK}, based on their scaling analysis, 
have proposed a phase diagram where the metallic phase can exist only in weak disorder 
regime under finite magnetic fields. In their phase diagram, a direct 
superconductor-insulator transition occurs in a relatively high disorder 
regime. While this discrepancy between the two phase diagrams is a puzzle 
that needs to be solved, we note that all the samples interpreted to 
exhibit a direct superconductor-insulator transition in Ref. \cite{SBK} 
are InO material systems which are known to have a wide resistive superconducting 
transition \cite{Shahar}. Typically, the width of the transition in InO 
($\Delta T_{c}/T_{c}$), with $\Delta T_{c}$ being the temperature interval corresponding to the resistance 
change from 90\% to 10\% of normal state resistance at B = 0, is larger 
than that observed in Ta systems by several times or more. The width of the 
resistive transition, $\Delta T_{c}/T_{c}$, can be a measure of 
disorder at a length scale longer than the superconducting coherence length. However, what role 
this long length scale disorder plays is not clear at present.

Finally, we turn to the discussion on the metallic phase appearing in the 
B = 0 finite temperature region [for example, the metallic phase along the 
temperature-axis in Fig.2(c)]. One might argue that the nonlinear 
transport with $d^{2}V/dI^{2} > 0$ of samples in this metallic phase could 
come from the Kosterlitz-Thouless (KT) mechanism \cite{KT}. The KT theory 
has been the framework to understand the onset of superconductivity in 2D at a 
finite temperature at B = 0. In this picture, the superconducting 
transition is described as a thermodynamic instability of 
vortex-antivortex pairs. The transport in the superconducting phase is 
expected to be nonlinear in a power law fashion, $V \propto I^\alpha$ with 
$\alpha > 3$, due to the current-induced vortex pair dissociations 
\cite{Halperin}. It has been argued \cite{MKM} that in a real system a 
strong finite size effect can induce free vortices altering the power law. The resulting nonlinear transport 
obtained in a numerical simulation \cite{MKM} closely resembles what is 
observed in our samples in their metallic phase. However, as reported 
earlier \cite{Yoon1, Yoon2}, we observe two transport properties that 
cannot be reconciled with the KT picture: One is the extraordinarily long relaxation time, up to several 
seconds, and the other is the development of electronic instabilities 
appearing as a sharp hysteresis in I-V curves at low temperatures. These 
observations indicate that the metallic nonlinear transport in the B = 0 
and finite temperature region is unlikely related to the KT mechanism, and 
raise the possibility that the KT transition might have been pre-emptied 
by a first order vortex pinning-depinning transition.

To summarize, we have mapped a phase diagram in T-B-disorder space for 
homogeneously disordered superconducting Ta films. The phase diagram 
includes four different phases: the superconducting phase identified by 
the presence of electronic instability possibly due to vortex 
pinning-depinning mechanism, the metallic phase by $d^{2}V/dI^{2} > 0$, 
the insulating phase by $d^{2}V/dI^{2} < 0$, and the normal conducting 
phase by $d^{2}V/dI^{2} = 0$. The resulting phase diagram shows that the 
superconducting phase is completely surrounded by the intervening metallic phase 
prohibiting a direct superconductor-insulator transition at any disorder, 
and that the metallic phase extends to B = 0 finite temperature region.

The authors acknowledge fruitful discussions with G. Refael and H. Fertig. 
This work was supported by the NSF through Grant No. DMR-0239450.

\bibliography{draft}

\end{document}